\documentclass[journal, twocolumn, 10pt, letterpaper]{IEEEtran}
\usepackage{graphics}
\usepackage{amsmath}
\usepackage{cite}
\usepackage{amsthm}
\usepackage{amssymb}
\usepackage{color}
\usepackage{trig}
\usepackage{tikz}

\newtheorem{definition}{Definition} [section]
\newtheorem{theorem}{Theorem}[section]
\newtheorem{lemma}{lemma}[section]

\newtheorem{example}{Example}[section]

\begin{document}

\title{Electricity Pooling Markets with Strategic Producers Possessing Asymmetric Information I: Elastic Demand}
\author{%
Mohammad Rasouli and Demosthenis Teneketzis, Fellow IEEE \\
Email: rasouli@umich.edu, teneket@umich.edu\\
Department of EECS\\
University of Michigan\\
Ann Arbor, MI , 48109-2122
}
\date{}

\maketitle

\begin{abstract}
\label{abstract}
In the restructured electricity industry, electricity pooling markets are an oligopoly with strategic producers possessing private information (private production cost function). We focus on pooling markets where aggregate demand is represented by a non-strategic agent. We consider demand to be elastic. 

We propose a market mechanism that has the following features. (F1) It is individually rational. (F2) It is budget balanced. (F3) It is price efficient, that is, at equilibrium the price of electricity is equal to the marginal cost of production.  (F4) The energy production profile corresponding to every non-zero Nash equilibrium of the game induced by the mechanism is a solution of the corresponding centralized problem where the objective is the maximization of the sum of the producers' and consumers' utilities.

We identify some open problems associated with our approach to electricity pooling markets.
\end{abstract}


\section{Introduction}
\label{Introduction}

\subsection{Motivation}

The electricity industry has been traditionally regulated by government to run under a predetermined-price cost-minimizing monopoly. In past decades, there has been world-wide tendency toward restructuring the industry toward a free competitive market \cite{Kagiannas}. By restructuring the industry, we will have oligopoly instead of monopoly. There have been different practices of electricity restructuring with California, Pennsylvania-new jersey-Maryland (PJM), British and Australian markets as some of the most prominent ones.

Electricity markets are the heart of the industry restructuring. Electricity as a trading commodity has unique features \cite{Wilson}. In particular, network flows are interconnected through KVL and KCL laws. So, trade of electricity between two nodes in the network affects the flows in the other nodes . Moreover, the flows are limited by physical capacity constraints of the lines. These limits have put the possibility of free bilateral trades in question and pooling markets have been proposed as a solution \cite{SioshansiOren}. In pooling markets, Independent System Operator (ISO) is a non-profit making entity that coordinates the flow of electricity in the network. ISO stands between consumers and producers and runs the market. He collects all the bids from the consumer and producer side and dispatches the electricity.

In general, there are three approaches to run pooling markets: Cournot, Bertrand and the supply function model \cite{Nanduri}. The first approach is based on the Cournot model. Here, each producer only bids his production amount. The consumers buy all the production but at a price that is a function of the total production. References \cite{Varaiya1}, \cite{Motto} and \cite{YaoOren} adopt this model for their studies and reference \cite{Ventosa} provides a comprehensive study of Cournot models in electricity markets. The British market originally, and the Australian National Electricity Market lie in this category \cite{SioshansiOren}.

The second approach is the Bertrand model. Here, every producer has a predetermined amount of generation and bids a price. The ISO then starts with the lowest price proposal and clears the market.

The third and most popular model is the supply function model (\cite{SioshansiOren}, \cite{Contreras}). Here, each producer sends a complete price-production curve to the ISO. This curve specifies the amount of production at every price level. ISO then runs an optimization to meet the demand with the lowest total cost. US electricity markets including California \cite{Borenstein2}, Pennsylvania New Jersey Maryland  (PJM) \cite{Ott} and Midwest ISO (MISO) \cite{Faruqui} are examples of markets with price-production bids. British market under reform is also in this category \cite{SioshansiOren}.

Reference \cite{Ventosa} presents a comprehensive survey of the literature on Bertrand vs. supply function auctions. Both of these markets are primarily based on multi-unit auctions where more than one unit of the same type is auctioned \cite{David}. Price-wise, these auctions can be in two categories, namely, uniform or discriminatory price auctions. Reference \cite{Nanduri} presents a comprehensive survey of the literature on uniform vs. discriminatory price auctions. In a uniform price auction, all selected suppliers are paid a uniform price, equal to the market clearing price. In a discriminatory auction, the suppliers are selected in a manner similar to the uniform auction, but are paid according to their own bids instead of the market clearing price.

Even though studies have been performed to identify the best auction format for electricity markets, conclusions as to the efficiency of the auction formats are still unclear \cite{Nanduri}. Because the analysis of electricity markets is very complex, so far electricity market equilibrium models either do not consider strategic bidding behavior or assume that players have all relevant information about the other players' characteristics and behavior \cite{Weidlich}. There is no theoretical background (analytical investigation) on the efficiency and performance of the currently proposed auctions.

The above discussion summarizes the state of the art in the design of electricity pooling markets. There are several open issues/problems associated with electricity pooling markets, such as the presence of strategic players with asymmetric information, inelasticity of demand in the market, the limits imposed by the transmission network and dynamic markets that run over a finite horizon. The study of electricity pooling markets with strategic players possessing asymmetric information for the case of elastic demand motivates the study presented in this paper.

This paper is a the first of a series of two papers. The second paper addresses electricity pooling markets where demand is inelastic.
\subsection{Contributions of the paper}

In this report we focus on a static pooling market with strategic producers possessing private information, non-strategic (price-taking) elastic demand, and no transmission constraints. We adopt Nash Equilibrium (NE) as a solution/equilibrium concept. The interpretation of NE is the same as in \cite{demos3}- \cite{sharma3}.

We assume demand to be elastic. We propose a market mechanism which has the following features. \textbf{(F1)} It is individually rational. That is, strategic producers voluntarily participate in the pooling market. \textbf{(F2)} It is budget balanced. That is, the mechanism does not create any budget surplus or any budget deficit. \textbf{(F3)} It is price efficient. That is, the demand is paying a price equal to the marginal cost of producing the next one unit of energy. \textbf{(F4)} The energy production profile corresponding to every non-zero NE of the game induced by the mechanism is a solution of the corresponding centralized problem, i.e. the problem the ISO would solve if it had access to the producers' private information. Furthermore, if zero production is the only NE of the game induced by the mechanism, the energy production profile corresponding to that is a solution of the corresponding centralized optimization problem.

The mechanisms/game forms presented in this report are distinctly different from currently existing mechanisms for electricity pooling markets. There are two key differences between our mechanism and currently existing mechanisms. 
\begin{enumerate}
\item In terms of the type of information exchange (the message space of the mechanism)
\item In terms of the performance 
\end{enumerate}
We elaborate on these key differences.

(1) In the Cournot model, agents report the amount of electricity they intend to supply to the pooling market. In the Bertrand model, agents report the price they intend to charge per unit of produced energy. In the supply function model, each agent reports its price-production curve. In our mechanism, each agent/producer reports the amount of energy it intends to produce along with the price per unit of energy it wishes to charge.

(2) In current studies, whenever performance analysis is done, there is no comparison between the performance achieved by the proposed market and the optimal centralized performance. We propose a market model where agents are strategic and possess private information. We present a complete analysis of the market and prove that it possesses features (F1)-(F4). 

We also note that, in many electricity markets proposed so far, strategic behavior or private information are not taken into account. Even when strategic behavior is taken into account, such as in supply-function models, strategic producers are assumed to report their true price-production curve to the ISO. In our opinion, this assumption is inconsistent with strategic behavior.

\subsection{Organization}
The rest of the paper is organized as following. The model of the market analyzed/studied in this paper is introduced in Section \ref{Model}. The objective is presented in section \ref{Objective}. The centralized optimization problem associated with the model of Section \ref{Model} and the objective of Section \ref{Objective} is presented in Section \ref{Centralized Problem}. The analysis of the problem with elastic demand appears in \ref{Elastic}. Discussion of open problems associated with our approach to electricity pooling markets appears in Section \ref{Conclusion}. The proofs of our results are presented in Appendix \ref{appendix_A}-\ref{appendix_B}. Examples illustrating our approach and results appear in Appendix \ref{appendix_D}.

\section{The Model}
\label{Model}
We consider a pooling market consisting of an ISO, $N$ producers, and consumers who are represented by their aggregate demand. Let $I=\{1,2,...,N\}$ denote the set of producers. We make the following assumptions:

(A1) The number of of producers, $N$, is fixed and common knowledge among the ISO, producers and consumers; furthermore, $N>3$.

(A2) Producers are strategic and self-profit maximizers.

(A3) Each producer $i$ has a fixed capacity $x_i > 0$, $i=1,2,...,N$, which is common knowledge among the producers and the ISO.

(A4) The cost function $C_i(.)$, $i=1, 2, ..., N$, of energy production is the producer $i$'s private information.  Also, $C_i(.)  \in \mathcal{C}_i$, where the function space $\mathcal{C}_i$ is common knowledge among producers and the ISO.

(A5) The functions $C_i(.)$, $i=1, 2, ..., N$, are convex; furthermore, for all $i$, $i=1,2, ..., N$, 
\begin{eqnarray}
\label{C_0}
C_i(0)=0,\\
\label {C_first_der}
C_i^{'}(e_i) > 0,\\
\label{C_second_der}
C_i^{''}(e_i) > 0,
\end{eqnarray}
for all $e_i>0$, where $e_i$ denotes the amount of energy produced by producer $i$, and $C_i^{'}(.)$ and $C_i^{''}(.)$ denote the first and second derivatives, respectively, of $C_i(.)$. 

(A6) Producer $i$'s utility function is
\begin{equation} 
\label{utility_consumers}
u_i(e_i, t_i)= -C_i(e_i)+t_i
\end{equation}
where $t_i$ denotes the amount of money producer $i$ receives for the energy it produces. 

(A7) The demand is elastic. It consumes the whole production.

(A8) The consumers' utility when they consume $d$ units of energy is $u_d(d)$; $u_d(.)$ is common knowledge among the ISO, producers and consumers.

(A9) The function $u_d(.)$ is concave with
\begin{eqnarray}
\label{u_d_0}
u_d(0)=0\\
\label {u_d_first_der}
u_d^{'}(d) > 0\\
\label{u_d_second_der}
u_d^{''}(d) < 0
\end{eqnarray}
for all $d>0$, where $u_d^{'}(.)$ and $u_d^{''}(.)$ denote the first and second derivatives, respectively, of $u_d(.)$.

(A10) The consumers' total utility is 
\begin{equation}
\label{utility_demand}
u_d(d)-\sum_{i\in I} t_i.
\end{equation}
where $d$ is the total energy consumed and $\sum_{i\in I} t_i$ denotes the amount of money demand pays to the producers for the energy consumed.

(A11) The ISO is a social welfare maximizer. From (\ref{utility_consumers}) and (\ref{utility_demand}) the social welfare function for the elastic demand is 
\begin{eqnarray}
\begin{aligned}
\label{social_welfare_elastic}
&W_1(e_1, e_2, ..., e_N) = u_d(\sum_{i\in I} e_i)-\sum_{i\in I} t_i-\sum_{i\in I} C_i(e_i)+\nonumber \\
&\sum_{i\in I} t_i  = u_d(\sum_{i\in I} e_i)-\sum_{i\in I} C_i(e_i). 
\end{aligned}
\end{eqnarray}

(A12) No transmission constraints are taken into account in the energy distribution. 

We now briefly discuss Assumptions (A1)-(A12).

Reference \cite{Varaiya1} argues that after restructuring, the market will change from a monopoly to an oligopoly. Therefore, we assume a finite $N >3$ number of producers.

Strategic self-profit maximizing behavior of the producers, (A2), is a mainstream assumption in industry restructuring. 

(A3) is based on the assumption that no producer invests on his generation capacity or retires any of his generation plants during the market time.

(A4) is the basis of the producers' strategic behavior. Cost functions are private information because they are dependent on variables which are not observable, including production technology, plant management, failures and resource limits \cite{Motto}. The assumptions on each $C_{i}(.)$, expressed in (\ref{C_0})-(\ref{C_second_der}) are standard in the literature (See \cite{Ventosa} and papers referenced in it).

The producers' utility functions, defined by (\ref{utility_consumers}) in (A6), are quasilinear. They include a cost of $C_i(e_i)$ which is then compensated by a payment $t_i$ from the consumers. 

The demand is elastic and consumes all of the productions (A7).

The strictly increasing and concave nature of $u_d (.)$, (A8), is a common assumption on the utility of the demand side \cite{Mascollel}. We assume ISO and producers have the same observation and estimation of the demand utility function. Therefore, $u_d(.)$ is common knowledge among the ISO and the producers.

Demand is required to pay a total of $\sum_{i\in I} t_i$ to the ISO to be distributed among producers. Therefore, the total utility of the consumers is as in (\ref{utility_demand}).

(A7) and (A10) imply that the demand is non-strategic. Demand does not bid in the market and does not decide on its amount of production and its payment. Reference \cite{elmaghrabyOren} has the same assumption in the pooling market and argues that it is consistent with most currently operating and proposed power auctions.

Considering the form of the producers' utilities as well as the consumers' utility, the non-profitmaker ISO aims to maximize the social welfare defined in Eq. (\ref{social_welfare_elastic}).

In this current model, transmission constraints are not taken into account (A12). Reference \cite{elmaghrabyOren} adopts same assumption for pooling markets and discusses that it is consistent with the UK system , the California power exchange, the Victoria pool and other systems around the world where transmission constraints and congestion management are handled outside the power auction. Transmission constraints have not been taken into account in other works that have also focused on the market side of electricity markets (See \cite{Varaiya1}, \cite{Ventosa} and the papers referenced therein).

\section{Objective and Method of Approach}
\label{Objective}

The ISO's objective is to maximize the social welfare function given by Eq. (\ref{social_welfare_elastic}) under the constraints imposed by (A1)-(A12) along with the requirement that the capacity constraints of the producers are satisfied.

To achieve this objective, we proceed as follows. We first consider the centralized optimization problem the ISO would solve if he had perfect knowledge of the cost functions, $C_i(.)$, $i=1,2,...,N$. The solution of this centralized problem would give the best possible performance the ISO can achieve.

Afterwards, we design a mechanism/game form that possesses properties (F1)-(F4). The above properties are obtained via the creation of a tax function which incentivizes each strategic producer to align his own individual objective with the social welfare. The specification and interpretation of the mechanism and its tax function appears in Section \ref{Elastic}.
\section{The Centralized Problem}
\label{Centralized Problem}

By Eq. (\ref{social_welfare_elastic}), the ISO's centralized problem is 
\begin{eqnarray}
\label{MAX1}
\max_{e_i , i\in I} & & u_{d}({\sum_{i\in I} e_i}) -\sum _ {i\in I}  C_i (e_i)\nonumber\\
s.t. & & 0 \leq e_i \leq x_i.
\end{eqnarray}
We call the above problem \textit{\textbf{MAX1}}. 

Assumptions (A3), (A5) and (A9) imply that in \textit{\textbf{MAX1}}, the objective function is strictly concave and the set of feasible solutions is non-empty, convex and compact. Therefore, \textit{\textbf{MAX1}} has a unique solution, and the Karush-Kuhn-Tucker (KKT) conditions are necessary and sufficient for optimality. The KKT conditions are useful for the analysis of the mechanism proposed in this paper. That is why they are presented in Appendix \ref{appendix_A}.

\section{The Mechanism for Elastic Demand}
\label{Elastic}
We first specify the mechanism, then we interpret its elements, mainly the tax function, and, finally, we study the properties of the mechanism. We illustrate the mechanism via two examples that appear in Appendix \ref{appendix_D}.

\subsection{Specification of the Mechanism}
\label{elastic_mechanism}

A game form/mechanism is described by $(\mathcal{M}, h)$, where $\mathcal{M}$ is the message/strategy space and $h: \mathcal{M} \rightarrow \mathcal{A}$ is the function from message space to the space $\mathcal{A}$ of allocations.

We consider the following mechanism. 

\textbf{\underline{Message space}} Let $\mathcal{M}$ be
\begin{equation}
\mathcal{M}:=(\mathcal{M}_1 \otimes \mathcal{M}_2 \otimes ... \otimes \mathcal{M}_N), 
\end{equation}
where $\mathcal{M}_i$ is producer $i$'s message space,
\begin{equation}
\mathcal{M}_i:= [0, x_i]\times \mathbb{R}_{+}, i\in I
\end{equation}
and $m_i\in \mathcal{M}_i$ is of the form 
\begin{equation}
m_i=(\hat{e}_i, p_i)
\end{equation}
where $\hat{e}_i$ denotes the amount of energy producer $i$ proposes to produce, and $p_i$ denotes the price producer $i$ proposes per unit of energy. Note that $\hat{e}_i$ is restricted by $0 \le \hat{e}_i \le x_i$ and $p_i$ is restricted by $p_i \ge 0$.

\textbf{\underline{Allocation Space}}  Let $\mathcal{A}$ be
\begin{equation}
\mathcal{A}:=(\mathcal{A}_1 \otimes \mathcal{A}_2 \otimes ... \otimes \mathcal{A}_N), 
\end{equation}
where $\mathcal{A}_i$ is producer $i$'s allocation space
\begin{equation}
\mathcal{A}_i:= [0, x_i]\times \mathbb{R}, i\in I,
\end{equation}
and $a_i\in \mathcal{A}_i$ is of the form 
\begin{equation}
a_i=(e_i, t_i),
\end{equation}
where ${e}_i$ denotes the amount of energy producer $i$ is scheduled to produce, and $t_i$ denotes the tax (respectively subsidy) producer $i$ should pay (respectively receive).

\underline{\textbf{Outcome function}} $h: \mathcal{M} \rightarrow \mathcal{A}$

For each $\textbf{m}:=(m_1, m_2, ...,m_N) \in \mathcal{M}$ we have 
\begin{equation}
h(\textbf{m})=({\textbf{e}}, \textbf{t})=({e}_1,...,{e}_N,{t}_1,...,{t}_N),
\end{equation}
where
\begin{eqnarray}
\label{mechanism1}
{e_i} &=& \hat{e_i} \\
\label{tax_form_elastic}
t_i &=& p_{i+1} e_i - (p_i-p_{i+1})^2 - p_{i+1}^2 \zeta^2 + \phi_i \qquad \\
\zeta &=& | D(\overline{p})-\sum_{i\in I} e_i| \\
\overline{p}&=&\frac{\sum_{i\in I} p_i}{N}\\
\label{F}
D(\overline{p}) &=& {(u_d^{'})}^{-1}(\overline{p})\\
\label{elastic_phi}
\phi_i &=& (p_{i+1}-p_{i+2})^2 \\
\label{P_N+1}
p_{N+1} &:=&p_{1}.
\end{eqnarray}
We proceed to interpret and analyze the properties of the proposed mechanism.

\subsection{Interpretation of the Mechanism}
Since the designer of the mechanism, i.e. ISO, can not alter the producers' cost functions, $C_i(.)$, $i=1,2,...,N$, even if he knew their functional form, the only way it can achieve his objective is through the use of appropriate tax incentives/tax functions. The tax incentive of our mechanism for produce $i$ consists of three components, that is, 
\begin{eqnarray}
t_i &=& t_{i,1} + t_{i,2} + t_{i,3},
\end{eqnarray}
where
\begin{eqnarray}
t_{i,1} &=& p_{i+1} e_i \\
t_{i,2}  &=& -(p_i-p_{i+1})^2 - p_{i+1}^2 \zeta^2 \\
t_{i,3}  &=& \phi_i.
\end{eqnarray}

The term $t_{i,1}$ specifies the amount user $i$ receives for its production $e_i$ from the demand side. It is important to note that the price per unit of electricity energy that a producer is paid is determined by the message/proposal of other producers. Thus, a user does not control the price per unit of electricity it provides.

The term $t_{i,2} $ provides the following incentives to the producers: (1) To bid/propose the same price per unit of produced energy (2) To collectively propose a total electricity supply that meets the optimal demand at the proposed price. The incentive provided to all users to bid the same price per unit of produced energy is described by the term $(p_i-p_{i+1})^2$, which is a positive punishment paid by producer $i$ for deviating from the price proposal of producer $i+1$.  The incentive provided to all producers to collectively propose a total production that meets the optimal demand is captured by the term $p_{i+1}^2 \zeta^2$.Note that $D(\overline{p})$ is the optimal demand for price $\overline{p}$, because it solves the optimization problem below:
\begin{equation}
\label{optimal_demand}
D(\overline{p})= \arg \max_{d \ge 0} u(d)-\overline{p}d.
\end{equation} 
$t_{i,2}$ can be thought of as the tax payments the ISO collects from the producers in order to align their productions with the social welfare maximizing production profiles.

The goal of $t_{i,3}$ is to lead to a balanced budget of payments by the producers to the ISO. After collecting $\sum_{i\in I} t_{i,2}$ of tax from producers, ISO distributes $ t_{i,3} $ among them in order to achieve budget balance, that is,
\begin{equation}
\sum_{i\in I} [t_{i,2}+t_{i,3}] = 0 .
\end{equation}
Note that, $t_{i,3}$ is not controlled by producer $i$'s messages, so $t_{i,3}$ does not affect producer $i$'s strategic behavior.

\subsection {Properties of the Mechanism}

The properties possessed by the proposed mechanism are described by Theorems (\ref{trivial_NE_elastic})-(\ref{budget_efficiency_elastic}) and lemma (\ref{lemma1_elastic}). The proof of all these properties are presented in Appendix \ref{appendix_B}.

\begin{theorem}
\label{trivial_NE_elastic}
(Existence of NE) One of the NE of the game induced by the proposed mechanism is $m_i^* = (0,0)$, for all $i\in I$. The corresponding production profile and taxes at this equilibrium are zero.
\end{theorem}

\begin{definition} We call the NE where for all $i\in I$, $m_i^*=(0, 0)$ a trivial Nash Equilibrium. We call any other NE of the game induced by the proposed mechanism a non-trivial NE.
\end{definition}

\begin{theorem} \label{feasibility_elastic} (FEASIBILITY) If $\textbf{m}^*= (\hat{\textbf{e}}^*, \textbf{p}^*) = (\hat{e}_1^*, \hat{e}_2^*, ..., \hat{e}_N^*, p_1^*, p_2^*, ... , p_N^*)$ is a non-trivial NE point of the game induced by the proposed mechanism, then the allocation $\textbf{e}^*$ is a feasible solution of problem \textbf{MAX1}, i.e.
\begin{eqnarray}
\label{feasibility_1_elastic}
\left. D(\overline{p})\right|_{m^*} - \sum_{i\in I} e_i^* = 0,\\
\label{zeta0_elastic}
\zeta^*=0.
\end{eqnarray}
\end{theorem}

\begin{lemma}
\label{lemma1_elastic}
Let $\textbf{m}^*$ be a non-trivial NE. Then for every producer $i \in I$ we have,
\begin{eqnarray}
\label{price_same_elastic}
p_i^*=p_{i+1}^*&=&p^*\\
\label{price_elastic_equal}
p^* &=& u^{'}(\sum_{i\in I} e_i^{*})\\
\label{tax_equ_elastic}
t_i^*&=&p^* e_i^*\\
\label{tax_der_equ_elastic}
\left.\frac{\partial t_i}{\partial e_i}\right|_{\textbf{m}^*} &=& p^*.
\end{eqnarray}
\end{lemma}

\begin{theorem} 
\label{Strong_Implementation_elastic} Consider a solution $\textbf{e}^*=(e_1^*, e_2^*, ..., e_N^*)$ of the centralized problem $\textbf{\textit{MAX1}}$. There exists a NE, $\textbf{m}^*= (\hat{e}_1^*, \hat{e}_2^*, ..., \hat{e}_N^*, p_1^*,p_2^*, ... ,p_N^*)$ of the game induced by the mechanism such that the production profile corresponding to $\textbf{m}^*$ is equal to $\textbf{e}^*$.
\end{theorem}

\begin{theorem} 
\label{Nash_Implementation_elastic}
(i) Consider any non-trivial NE $\textbf{m}^*$ of the game induced by the mechanism. Then, the production profile $\textbf{e}^*$ corresponding to $\textbf{m}^*$ is an optimal solution of the centralized problem $\textbf{MAX1}$.  (ii) If trivial NE is the only NE of the game induced by the mechanism, then the zero production profile corresponding to that is the optimal solution of the centralized problem $\textbf{MAX1}$.
\end{theorem}

\begin{theorem}
\label{Individual_Rationality_elastic} (INDIVIDUAL RATIONALITY) The proposed game form is individually rational, that is at every NE of the game induced by the mechanism, the corresponding allocation $(\textbf{e}^*, \textbf{t}^*)$ is weakly preferred by all users to the initial allocation $(\textbf{0} , \textbf{0})$.
\end{theorem}

Theorems (\ref{Strong_Implementation_elastic}),  (\ref{Nash_Implementation_elastic}) and (\ref{Individual_Rationality_elastic}) show the game induced by the mechanism has a set of NE including all the solutions to the centralized problem plus a trivial NE of all zero prices and all zero productions. The trivial NE is Pareto dominated by any other NE of the game. Note that for high cost of production, the centralized problem may have only corner solutions of $e_i^*=0, \quad \forall i\in I$. Then this corner solution corresponds to the trivial NE outcome of the game, which is the unique NE in this case.

\begin{theorem}
\label{budget_balance_elastic}
(BUDGET BALANCE) The mechanism is budget balanced both at equilibrium and off equilibrium, that is the payments from all producer and consumers sum up to zero.
\end{theorem}

\begin{theorem} 
\label{budget_efficiency_elastic}
(PRICE EFFICIENCY) The mechanism is price efficient; that is at equilibrium, the demand is paying a price equal to the marginal utility of the next one unit of production.
\end{theorem}

The result of Theorem (\ref{budget_efficiency_elastic}) shows that the game induced by the proposed mechanism incentivizes the producers to reveal the true marginal cost of production of the system at equilibrium.

\section{Conclusion and Reflections}
\label{Conclusion}

Electricity restructuring has changed the industry from a monopoly into an oligopoly where energy producers are strategic players with private information and market power. In this new environment, electricity can be traded through bilateral contracts or pooling markets. In this paper we focused on pooling markets with strategic producers possessing private information, non-strategic consumers with elastic demand, and no transmission constraints. We designed a mechanism which has properties (F1)-(F4). Achieving these features all together distinguishes our mechanism from any other market design available in the literature. In the games induced by the proposed mechanism, the only NE besides the one that results in a production profile equal to the centralized solution is the trivial one which is Pareto dominated by the centralized solution NE. It is worth noting that price efficiency is achieved even though customers (represented by their aggregate demand) are not strategic.

The game form presented in this paper ensures that the desired allocations
are achieved at equilibria without specifying how an equilibrium is reached.
That is, this game form/mechanism does not include an iterative
process that determines how the NE of the game induced by the mechanism are
computed by the users. The lack of such iterative processes for decentralized
resource allocation problems where strategic users possess private information
is a major open problem in mechanism design. The major difficulty in
constructing iterative algorithms that guarantee convergence to NE is the following. Consider an algorithm/iterative process for a decentralized allocation problem
where strategic users possess private information. At each stage of the algorithm, each user updates its message. After an update, a user, say user $i$, can report any message it deems beneficial to itself, and other users may not be able to check whether or not user $i$ is following the rules of the algorithm. Consequently, the construction of the iterative process must provide incentives to the users to follow the rules at each stage of the algorithm. Such a provision of incentives must be based, in general, on all the information available at the current stage and must, in general, take the whole future into account. Algorithms with the above features are currently unavailable \cite{demos3}.

Future problems along this line of research include the consideration of inelasticity for demand, transmission constraints, and dynamic games for a number of markets over a time horizon.

\textbf{Acknowledgment}: This research was supported in part by NSF Grant CNS-1238962. The authors thank Hamidreza Tavafoghi for many useful discussions.
\begin{appendices}
\section{\textbf{The Karush-Kuhn-Tacker condition for problem \textit{MAX1}}}
\label{appendix_A}
The Lagrangian for \textit{\textbf{MAX1}} is
\begin{eqnarray}
\label{Lagrangian_MAX1}
\mathcal{L}_{MAX1} &=& u_{d}(\sum_{i\in I} e_i) - \sum_{i\in I} C_i(e_i) \nonumber\\
&&+\sum_{i\in I}\mu_i (x_i-e_i)+\sum_{i\in I}\nu_i e_i,
\end{eqnarray}
and KKT conditions are, $\forall i\in I$,
\begin{eqnarray}
\label{KKT}
\label{KKT_MAX1_first}
\left. \frac{\partial u_{d} (\sum_{i\in I} e_i)}{\partial e_i} \right|_{\textbf{e}^*} - \left. \frac{\partial C_i}{\partial e_i} \right|_{\textbf{e}^*} - \mu_i^* + \nu_i^* &=& 0 ,\\ 
{\mu}^*_i(x_i-e_i^*)&=&0,\\
{\nu}^*_i e_i^*=0,\quad {\nu}_i^* \geq 0,\quad \label{KKT_MAX1_last} {\mu}_i^* &\geq&0.
\end{eqnarray}

\section{\textbf{Proof of Theorems (\ref{trivial_NE_elastic})-(\ref{budget_efficiency_elastic}) and of lemma (\ref{lemma1_elastic})}}
\label{appendix_B}

\textbf{Proof of Theorem \ref{trivial_NE_elastic}} 
Consider producer $i$ and let $m_{j}^* = (0,0) \quad \forall j\neq i$. The first order conditions of producer $i$'s best responses are
\begin{eqnarray}
\label{Trivial_FOC_p_elastic}
\left. \frac{\partial u_i}{\partial p_i}\right|_{\textbf{m}_{-i}^*} &=& -2(p_i-p^*_{i+1}) = -2 p_i\nonumber\\
 \label{Trivial_FOC_e_elastic}
\frac{\partial u_i}{\partial e_i}|_{\textbf{m}_{-i}^*} &=& -C^{'}_{i}(e_i)+p_{i+1}-p_{i+1}^2 \left.\frac{\partial \zeta^2}{\partial e_i}\right|_{\textbf{m}_{-i}^*}= -C_i^{'}(e_i).\nonumber
\end{eqnarray}
Considering Eqs (\ref{C_0})-(\ref{C_first_der}) and solving for first order conditions, we have $p_i^*=0,\quad e_i^*=0$. As a result, no producer has the incentive to deviate from $(0,0)$ and therefore, $\textbf{m}^*=\{ m_i^* = (0,0), \forall  i\in I\}$ is a NE For this set of messages, Eq. (\ref{mechanism1}), (\ref{tax_form_elastic}) and (\ref{elastic_phi}) result that $\textbf{e}^*=\textbf{0}$ and $t_i^*=0$.

\textbf{Proof of Theorem \ref{feasibility_elastic}}
Since $\phi_i$ does not depend on producer $i$'s message
\begin{equation}
\label{NE_der_p}
\frac{\partial \phi_i}{\partial p_i}  = \frac{\partial \phi_i}{\partial e_i}  = 0.
\end{equation} 

Consider two cases.

\underline{\textit{Case 1:}} For all $i \in I$, $p_i^* >0$.

Here, $p_i^* \geq 0$ is not binding. Therefore, at NE, for all $i \in I$,
\begin{equation}
\label{NE_der_p_elastic}
\left.\frac{\partial u_i}{\partial p_i}\right|_{\textbf{m}^*} =\left.\frac{\partial t_i}{\partial p_i}\right|_{\textbf{m}^*} = -2(p^*_i-p^*_{i+1})-2 {p_{i+1}^*}^2 {\zeta^*} \left.\frac{\partial \zeta}{\partial p_i}\right|_{\textbf{m}^*}= 0. \nonumber
\end{equation} 
Summing up over all $i \in I$ we get:
\begin{eqnarray}
\label{Sum_der_p_elastic}
\sum _{i\in I} \left.\frac{\partial t_i}{\partial p_i}\right|_{\textbf{m}^*} &=&-\sum_{i\in I} [ 2(p^*_i-p^*_{i+1})+2 {p_{i+1}^*}^2 {\zeta^*} \frac{\partial \zeta}{\partial p_i}|_{\textbf{m}^*}] \nonumber\\
&=& -2 (\sum_{i \in I} {p_{i+1}^*}^2)  {\zeta^*} \left.\frac{\partial \zeta}{\partial p_i}\right|_{\textbf{m}^*} = 0.
\end{eqnarray}
In addition,
\begin{equation}
\label{zeta_der_elastic}
\left.\frac{\partial \zeta}{\partial p_i}\right|_{\textbf{m}^*} = \left.\frac{\partial \zeta}{\partial p_j}\right|_{\textbf{m}^*} = sign(D(\overline{p})-\sum_{i\in I} e_i^*) \frac{D^{'}(\overline{p})}{N};
\end{equation}
furthermore, from Eq. (\ref{F}),
\begin{equation}
\label{F_der_elastic}
D^{'}(\overline{p}) = \frac{-1}{u_{d}'' (\sum_{i\in I}e_i^*)}.
\end{equation}
Eqs. (\ref{u_d_second_der}), (\ref{Sum_der_p_elastic}), (\ref{zeta_der_elastic}) and (\ref{F_der_elastic}) imply that $\zeta^*=0$ or $[D(\overline{p})|_{m^*} - \sum_{i\in I} e_i^*] = 0$.

\underline{\textit{Case 2:}} There exists $i \in I$ $s.t.$ $p_i^*=0$.

We prove that in this case we have a trivial NE. 

Since $p_i \geq 0$ is binding at NE, we have 
\begin{equation}
\label{tax_p0_pos_elastic}
\left.\frac{\partial u_i}{\partial p_i}\right|_{\textbf{m}^*} \leq 0.
\end{equation}
Furthermore, 
\begin{eqnarray}
\label{tax_der_p0_elastic}
\left.\frac{\partial u_i}{\partial p_i}\right|_{\textbf{m}^*} &=& \left.\frac{\partial t_i}{\partial p_i}\right|_{\textbf{m}^*}= -2(p^*_i-p^*_{i+1})-2 {p^*_{i+1}}^2 {\zeta^*}\left.\frac{\partial \zeta}{\partial p_i}\right|_{\textbf{m}^*} \nonumber\\
&=&  2p^*_{i+1}-2 {p^*_{i+1}}^2 {\zeta^*}\left.\frac{\partial \zeta}{\partial p_i}\right|_{\textbf{m}^*}. 
\end{eqnarray}
Now assume $p_{i+1}^* > 0$. Then, from (\ref{zeta_der_elastic}), (\ref{tax_p0_pos_elastic}) and (\ref{tax_der_p0_elastic})
\begin{equation}
\label{zeta_zeta_der_elastic}
\left.{\zeta^*} \frac{\partial \zeta}{\partial p_i}\right|_{\textbf{m}^*} = {\zeta^*} \left.\frac{\partial \zeta}{\partial p_j}\right|_{\textbf{m}^*} > 0.
\end{equation}
Also, since $p_{i+1} \geq 0$ is not binding, 
\begin{eqnarray}
\label{p_i_1_not_binding_elastic}
\left.\frac{\partial t_{i+1}}{\partial p_{i+1}}\right|_{\textbf{m}^*} = -2(p^*_{i+1}-p^*_{i+2})-2 {{p_{i+1}^*}^2 }{\zeta^*} \left.\frac{\partial \zeta}{\partial p_i}\right|_{\textbf{m}^*}=0.
\end{eqnarray}
From (\ref{zeta_zeta_der_elastic}) and (\ref{p_i_1_not_binding_elastic}) it follows that
\begin{eqnarray}
p_{i+2}^* = p^*_{i+1}+{p_{i+1}^{*}}^{2}{\zeta^*} \left.\frac{\partial \zeta}{\partial p_i}\right|_{\textbf{m}^*} > 0.
\end{eqnarray}
Following the same argument, we obtain
\begin{eqnarray}
p_{i}^* = p^*_{i+1}+(\sum_{j\in I, j\neq i}  {p^*_{j+1}}^2) {\zeta^*} \left.\frac{\partial \zeta}{\partial p_i}\right|_{\textbf{m}^*}>0.
\end{eqnarray}
This contradicts the assumption of $p_i^*=0$. As a result, we should have $p_{i+1}^*=0$. Repeating the above argument we obtain $p_j^*=0, \quad \forall j \in I$. 

Next, we show that $e_i^*=0$ is the best response of producer $i$ to any $\textbf{m}_{-i}=\{m_j=(e_j, 0), \forall j\neq i\}$. We consider first order conditions for producer $i$'s best response.
\begin{eqnarray}
\left.\frac{\partial u_i}{\partial e_i}\right|_{\mathbf{m}_{-i}^*}&=& -C^{'} (e_i) + {p^*_{i+1}} - [2 {p^*_{i+1}}^2 |F(0)- e_i - \sum_{j\in I, j \neq i} e_j^* | \nonumber \\ 
&&\times  sign(F(0)- e_i - \sum_{j\in I, j \neq i} e_j^*)] = - C^{'}(e_i).
\end{eqnarray}
From (\ref{C_first_der}), this derivative is always negative; hence, $e_i^*=0, \forall i\in I$. The result is a set of trivial NE of all zero prices and zero products.

\textbf{Proof of Lemma \ref{lemma1_elastic}}
Since for all $i \in I$, $p_i^* > 0$, using Eq. (\ref{zeta0_elastic}) in Eq. (\ref{NE_der_p_elastic}) we obtain
\begin{equation}
\frac{\partial t_i}{\partial p_i} |_{\textbf{m}^*} = -2(p^*_i-p^*_{i+1})= 0.
\end{equation}
Therefore, 
\begin{equation}
\label{price_same_2_elastic}
p^*_i=p^*_{i+1}=p* \qquad \forall i\in I.
\end{equation}

$p^*=u^{'}(\sum_{i\in I} e_i^*)$ is a direct consequence of $\zeta^{*}=0$ and (\ref{price_same_2_elastic}). And one immediate consequence of Eqs (\ref{elastic_phi}) and (\ref{price_same_2_elastic}) is
\begin{equation}
\label{phi_0_elastic}
\phi_i^*=0 \qquad \forall i\in I.
\end{equation}
Finally, Eqs. (\ref{tax_form_elastic}), (\ref{feasibility_1_elastic}), (\ref{zeta0_elastic}), (\ref{price_same_2_elastic}) and (\ref{phi_0_elastic}) imply $t_i^* = p^* e_i^*$, and because $\zeta^*=0$,
\begin{equation}
\left.\frac{\partial t_i}{\partial e_i}\right|_{\textbf{m}^*} = p^* - 2 {p^*}^2 \zeta^* \left.\frac{\partial \zeta}{\partial e_i}\right|_{\textbf{m}^*} = p^*.
\end{equation}

\textbf{Proof of Theorem \ref{Strong_Implementation_elastic}}
We consider two cases.

\underline{\textit{Case 1:}} Interior solution ($e_i^*>0$ for some $i\in I$). Set, $\forall i\in I$,
\begin{eqnarray}
\begin{aligned}
\label{KKT_equivalency_2_interior_elastic}
&p_i^*&=&u_d^{'}(\sum_{i \in I}e_i^{*}),\quad  \hat{\mu}_i^*=\mu_i^*,\nonumber\\
&\hat{\nu}_i^*&=&{\nu}_i^*,\quad \hat{\theta}_i^*=0.\nonumber
\end{aligned}
\end{eqnarray}
Eqs. (\ref{KKT_MAX1_first})-(\ref{KKT_MAX1_last}) show that Eqs. (\ref{KKT_pro_first_elastic})-(\ref{KKT_pro_last_elastic}) are satisfied under the above selection.

\underline{\textit{Case 2:}} Corner solution ($e_i^*=0, \forall i\in I$). Set, $\quad \forall i\in I$,
\begin{eqnarray}
\begin{aligned}
\label{KKT_equivalency_2_corner_elastic}
&p_i^{*}=0,\quad \hat{\mu}_i^*=\mu_i^*=0,\nonumber\\
&\hat{\nu}_i^*= -\left.\frac{\partial C_i}{\partial e_i}\right|_{e_i=0},\quad \hat{\theta}_i^*=0.\nonumber
\end{aligned}
\end{eqnarray}
Eqs. (\ref{KKT_MAX1_first})-(\ref{KKT_MAX1_last}) show that Eqs. (\ref{KKT_pro_first_elastic})-(\ref{KKT_pro_last_elastic}) are satisfied under the above selection for corner solution.

\textbf{Proof of Theorem \ref{Nash_Implementation_elastic}}
(i) Let $\textbf{m}^*=(\hat{\textbf{e}}^*, \textbf{p}^*)=(\hat{e}_1^*, ...,\hat{e}_N^*, p_1^*, ..., p_N^*)$ be a NE of the game induced by the proposed mechanism. Then, $(\hat{e}_i^*, p_i^*)$ is a solution to producer $i$'s profit maximization problem, that is, 
\begin{eqnarray}
\label{Producer_Profit_elastic}
(\hat{e}_i^*, p_i^*)=\arg\max_{\hat{e}_i, p_i} & &- C_i(\hat{e}_i)+t_i\nonumber\\
\label{pro_capacity_constraint_elastic}
s.t. & & 0 \leq \hat{e}_i \leq x_i \nonumber\\
& & p_i \geq 0.
\end{eqnarray}
Call this problem \textbf{\textit{MAX2}}. The Lagrangian for this problem is
\begin{eqnarray}
\label{Lagrangian_ind_elastic}
\mathcal{L} _{MAX2}&=& - C_i(\hat{e}_i)+t_i+\hat{\mu}_i (x_i-\hat{e}_i)\nonumber\\
&+&\hat{\nu}_i \hat{e}_i+\hat{\theta}_i p_i
\end{eqnarray}
and the corresponding KKT conditions are
\begin{eqnarray}
\label{KKT_pro_first_elastic}
-\left.\frac{\partial C_i}{\partial \hat{e}_i}\right|_{\textbf{m}^*}+\left.\frac{\partial t_i}{\partial \hat{e}_i}\right|_{\textbf{m}^*}-\hat{\mu}_i^*+\hat{\nu}_i^* &=& 0,\\
\frac{\partial t_i}{\partial p_i} |_{\textbf{m}^*} - \hat{\theta}_i^* &=& 0,\\
\hat{\mu}^*_i(\hat{e}_i^*-x_i)=0,\quad \hat{\nu}^*_i \hat{e}_i^*=0,\quad \hat{\theta}^*_i p_i^*&=&0,\\
\hat{\nu}_i^* \geq 0,\quad \hat{\mu}_i^* \geq 0,\quad \label{KKT_pro_last_elastic},\hat{\theta}_i^* &\geq& 0.
\end{eqnarray}

To show that the allocations $e_i^*=\hat{e}_i^*$, $i\in I$, corresponding to $\textbf{m}^*$ are solutions of the centralized problem \textit{\textbf{MAX1}}, we construct the KKT parameters of the centralized problem based on the producers' profit maximization KKT parameters as $\mu_i^*=\hat{\mu}_i^*,\quad \nu_i^*=\hat{\nu}_i^*\quad \forall i\in I $.

Then, Eqs. (\ref{KKT_pro_first_elastic})-(\ref{KKT_pro_last_elastic}) along with lemma \ref{lemma1_elastic} show that Eqs. (\ref{KKT_MAX1_first})-(\ref{KKT_MAX1_last}) are satisfied. Consequently, $\hat{e}^{*}$, is a solution of the centralized problem \textit{\textbf{MAX1}}. The assertion of Theorem (\ref{Nash_Implementation_elastic}) follows since $\textbf{m}^*$ is an arbitrary NE of the game induced by the mechanism.

(ii) From Theorem (\ref{Strong_Implementation_elastic}), if there is only a unique NE, it should be corresponding to the unique solution of the centralized problem $\textbf{\textit{MAX1}}$.

\textbf{Proof of Theorem \ref{Individual_Rationality_elastic}}
Consider 3 cases.

\underline{\textit{Case 1:}} $e_i^*=0$. Then Eq. (\ref{tax_equ_elastic}) results in
\begin{equation}
u_i (\textbf{e}^*, \textbf{t}^*) = -C_i(e_i^*)-t_i^*= -C_i(0)-p_i^*\times 0 = 0.
\end{equation}
Therefore, the NE outcome is weakly preferred to the initial allocation.

\underline{\textit{Case 2:}} $ 0 < e_i^* < x_i $. The constraint in Eq. (\ref{pro_capacity_constraint_elastic}) is not binding and therefore the corresponding $\hat{\mu}_i^*$ and $\hat{\nu}_i^*$ are $0$. Then, Eqs. (\ref{tax_equ_elastic}), (\ref{tax_der_equ_elastic}) and (\ref{KKT_pro_first_elastic}) along with $\hat{\mu}_i^*=\hat{\nu}_i^*=0$ result in 
\begin{equation}
\label{Utility at equ 2_elastic}
u_i(\textbf{e}^*, \textbf{t}^*) =  -C_i(e_i^*)+t_i^* = -C_i(e_i^*) + C_i^{'}(e_i^*) e_i^*.
\end{equation}
Furthermore, from (\ref{C_0})-(\ref{C_second_der}), for the convex and increasing function $C_i$, 
\begin{equation}
\label{C_less_elastic}
C_i(e_i) < C_i^{'}(e_i) e_i \qquad \text{for any}\quad e_i>0.
\end{equation}
Combining (\ref{Utility at equ 2_elastic}) and (\ref{C_less_elastic}) we get, 
\begin{equation}
u_i(\textbf{e}^*, \textbf{t}^*) = -C_i(e_i^*) + C_i^{'}(e_i^*) e_i^* >0.
\end{equation}

\underline{\textit{Case 3:}} $ e_i^* = x_i $. Since the constraint $e_i \leq x_i$ is binding, $\hat{\mu}_i^*>0$, $\hat{\nu}_i^*=0$, and Eqs. (\ref{C_0}), (\ref{C_first_der}), (\ref{tax_der_equ_elastic}) and (\ref{KKT_pro_first_elastic}) imply
\begin{equation}
\label{price_upper_const_elastic}
p^*= \frac{\partial t_i}{\partial e_i} = C_i^{'}(e_i^*)+\hat{\mu}^* > C_i^{'}(e_i^*).
\end{equation}
Inequality (\ref{price_upper_const_elastic}) along with (\ref{tax_equ_elastic}) and (\ref{C_less_elastic}) result in
\begin{equation}
\begin{aligned}
&u_i(\textbf{e}^*, \textbf{t}^*)  -C_i(e_i^*)+t_i^*=\nonumber\\
&-C_{i}(e_i^*)+p^* e_i^* > -C_i(e_i^*) + C_i^{'}(e_i^*) e_i^* > 0.
\end{aligned}
\end{equation}

\textbf{Proof of Theorem \ref{budget_balance_elastic}}
Producer $i$ receives $t_i$ and demand pays $\sum_{i\in I }t_i$. Therefore, the sum of all payments adds up to zero at every message proposal.

\textbf{Proof of Theorem \ref{budget_efficiency_elastic}}
First consider non-trivial NE with producer $i \in I$ for which the production capacity constraints are not binding, i.e. $0 < e_i^* < x_i$. Therefore, $\hat{\mu}_i^*=\hat{\nu}_i^*=0$ in Eq. (\ref{KKT_pro_first_elastic}). This along with Eq. (\ref{mechanism1}) and (\ref{tax_der_equ_elastic}) imply that $ p^* = \left.\frac{\partial C_i}{\partial e_i}\right|_{e_i^*}$.
Next, consider the case of zero production. Here, by Theorem (\ref{trivial_NE_elastic}), demand pays tax equal to zero for zero production. Therefore, the trivial price can be set equal to the marginal cost of production.
\section{\textbf{Examples of Electricity Pooling Market with Elastic Demand}}
\label{appendix_D}

Here, we provide two examples of electricity pooling markets. At equilibrium, the first one results in positive production and the second one results in zero production. 

As we proposed in Section \ref{Conclusion}, we currently don't have an algorithm for computing the NE of the game induced by the mechanism. Nevertheless, we have proven that the production profiles corresponding to all non-trivial NE of the game induced by the mechanism are optimal solutions of the corresponding centralized problems. Thus, we obtain these production profiles as the solution of the centralized problem.
\begin{example}
\label{example_1}

This example illustrates the case where the game induced by the mechanism proposed in Section \ref{Elastic} has a non-trivial NE and the production profile corresponding to this NE is positive.

Consider a network of four producers with following cost functions and capacities
\begin{eqnarray}
\begin{aligned}
&C_1(e_1)=2e_1+e_1^2 \quad C_2(e_2)=3e_2+e_2^3\nonumber\\
&C_3(e_3)=4e_3+e_3^4\quad C_4(e_4)=5e_4+e_4^2\nonumber\\
&x_1=x_2=x_3=x_4=2.
\end{aligned}
\end{eqnarray}
The demand is elastic with utility function 
\begin{equation}
u_d(d)=40\sqrt{d}.
\end{equation}

The social welfare function, Eq. (\ref{social_welfare_elastic}) is
\begin{eqnarray} 
W(e_1, e_2, ..., e_4, d)= u_d(d)-\sum_{i=1,2,...,4} C_i(e_i) =\nonumber\\
40\sqrt{d} - [2e_1+e_1^2+3e_2+e_2^3+4e_3+e_3^4+5e_4+e_4^2].
\end{eqnarray}
and the corresponding problem \textbf{\textit{MAX1}} is 
\begin{align*}
\max_{e_i , i\in I}& 40\sqrt{\sum_{i\in I} e_i} - [2e_1+e_1^2+3e_2+e_2^3+4e_3+e_3^4+5e_4+e_4^2]\nonumber\\
&s.t. \quad 0 \leq e_i \leq x_i \nonumber
\end{align*}
The optimal production profile is
\begin{eqnarray}
e^*_1=2\quad e^*_2=1.5\quad e_3^*=1.1\quad e_4^*=0\nonumber
\end{eqnarray}


The market game in this problem follows from mechanism $(\mathcal{M}, h)$ of Section (\ref{Elastic}) with these specifications. \textbf{Message space} is $\mathcal{M}_i:= [0, 2]\times \mathbb{R}_{+}, i\in \{1,2,3,4\}$ where $m_i\in \mathcal{M}_i$ is of the form $m_i=(\hat{e}_i, p_i)$. \textbf{Allocation Space} is $\mathcal{A}:=(\mathcal{A}_1 \otimes \mathcal{A}_2 \otimes \mathcal{A}_3 \otimes \mathcal{A}_4),$ where $\mathcal{A}_i:= [0, 2]\times \mathbb{R}, i\in I,$ is producer $i$'s allocation space. And finally, \textbf{outcome function}, $h: \mathcal{M} \rightarrow \mathcal{A}$, for each $m:=(m_1, m_2,m_3,m_4) \in \mathcal{M}$ is defined by Eqs (\ref{mechanism1})-(\ref{P_N+1}). Note that here, $\overline{p}=\frac{\sum_{i\in I} p_i}{4}$ and $D(\overline{p}) = {(u_d^{'})}^{-1}(\overline{p})=p^2/1600$.

This game has a non-trivial NE of the form $\hat{e}^*_i=e_i^*, p_i^*=p^*$, where $p^*$ will be the marginal cost of production. By Eqs (\ref{price_elastic_equal}) and (\ref{tax_equ_elastic}) the price and the tax payments at equilibrium will be the following.
\begin{eqnarray}
\begin{aligned}
&p^*=\left. \frac{\partial C_2}{\partial e_2}\right|_{e^*}=\left. \frac{\partial C_3}{\partial e_3}\right|_{e^*}=9.35 \$/MWH \nonumber\\
&t_1^*=18.7\quad t_2^*=13.6\quad t_3^*=10.3\quad t_4^*=0\nonumber
\end{aligned}
\end{eqnarray}

From Eqs. (\ref{utility_consumers}) and (\ref{utility_demand}), the utility of producers and the demand at equilibrium is the following.
\begin{eqnarray}
u_1=10.7, u_2=6.2, u_3=4.45, u_4=0, u_d-\sum_{i\in I} t_i=43.2\nonumber
\end{eqnarray}
Since all the producers utilities are positive, the game is individually rational. Price efficiency and  budget balance are also satisfied.  

Note that the game induced by the mechanism has also a trivial NE which is $m_i^*=(0,0)$ for all $i\in I$. At this no-trade situation, the utility of all producers as well as the demand will be zero; therefore, it is Pareto dominated by the non-trivial NE.
\end{example}
\begin{example}
This example illustrates the case where the game induced by the mechanism proposed in Section \ref{Elastic} has only a trivial NE and the production profile corresponding to this NE is all $0$.

Consider the same network as in Example (\ref{example_1}). Assume the demand is elastic with the following utility function:
\begin{equation}
u_d(d)= \left\{
   \begin{array}{l c l}
      100-(d-10)^2 &\qquad& 0 \leq d \leq 10\\
      100 &\qquad& otherwise
    \end{array}
\right.\nonumber
\end{equation}
Replacing $u_d$ in social welfare maximization, Eq. (\ref{MAX1}), results in the optimal production of $0$. The  game induced by the mechanism designed for elastic demand has only one equilibrium with $m_i^*=(0,0), \forall i\in I$ . This is a trivial equilibrium. Theorem (\ref{trivial_NE_elastic}) shows that the equilibrium price and the tax payments are all zero. Form Eqs. (\ref{utility_consumers}) and (\ref{utility_demand}), the utility of producers and the demand at equilibrium is also zero. Also, the mechanism is individually rational and budget balanced.
\end{example}
\end{appendices}
\bibliographystyle{IEEEtran}
\bibliography{IEEEabrv,Elec_Pooling_Markets_Elastic,collection}
\end{document}